\documentclass{article}
\usepackage{spconf,amsmath,graphicx}
\usepackage{amsfonts}
\usepackage{booktabs}
\usepackage{multirow}
\usepackage{hyperref}
\usepackage{siunitx}
\usepackage{mathtools}
\usepackage[subrefformat=parens]{subcaption}
\usepackage{comment}
\usepackage[linesnumbered,ruled,vlined]{algorithm2e}
\usepackage{color}

\def\equationautorefname~#1\null{(#1\null)}

\newcommand{\numcircle}[1]{\textcircled{\raisebox{-.3pt}{{\scriptsize #1}}}}
\let\oldnl\nl
\newcommand{\nonl}{\renewcommand{\nl}{\let\nl\oldnl}}
\newlength\mylen
\newcommand\myinput[1]{%
  \settowidth\mylen{\KwIn{}}%
  \setlength\hangindent{\mylen}%
  \nonl\hspace*{\mylen}#1}

\SetCommentSty{mycommfont}

\usepackage{xspace}
\makeatletter
\DeclareRobustCommand\onedot{\futurelet\@let@token\@onedot}
\def\@onedot{\ifx\@let@token.\else.\null\fi\xspace}
\def\eg{\emph{e.g}\onedot} 
\def\ie{\emph{i.e}\onedot}

\def\etal{\emph{et al}\onedot}
\makeatother

\newcommand{\vect}[1]{\mbox{\boldmath $#1$}}

\newcommand{\norm}[1]{\left\lVert#1\right\rVert}
\newcommand{\herm}[1]{#1^\mathsf{H}}

\newcommand{\inv}[1]{#1^{-1}}
\newcommand{\trace}[1]{\mathrm{tr}\left(#1\right)}

\def\appendixautorefname~#1\null{~#1 \null}
\newcommand{\relmid}{\mathrel{}\middle|\mathrel{}}

\makeatletter
\newcommand{\figcaption}[1]{\def\@captype{figure}\caption{#1}}
\newcommand{\tblcaption}[1]{\def\@captype{table}\caption{#1}}
\makeatother

\title{BLOCK-ONLINE GUIDED SOURCE SEPARATION}
\name{Shota Horiguchi \qquad Yusuke Fujita \qquad Kenji Nagamatsu}
\address{Hitachi, Ltd., Japan}
\begin{document}
\ninept

\maketitle
\begin{abstract}
We propose a block-online algorithm of guided source separation (GSS). GSS is a speech separation method that uses diarization information to update parameters of the generative model of observation signals. Previous studies have shown that GSS performs well in multi-talker scenarios.
However, it requires a large amount of calculation time, which is an obstacle to the deployment of online applications.
It is also a problem that the offline GSS is an utterance-wise algorithm so that it produces latency according to the length of the utterance.
With the proposed algorithm, block-wise input samples and corresponding time annotations are concatenated with those in the preceding context and used to update the parameters.
Using the context enables the algorithm to estimate time-frequency masks accurately only from one iteration of optimization for each block, and its latency does not depend on the utterance length but predetermined block length.
It also reduces calculation cost by updating only the parameters of active speakers in each block and its context.
Evaluation on the CHiME-6 corpus and a meeting corpus showed that the proposed algorithm achieved almost the same performance as the conventional offline GSS algorithm but with 32x faster calculation, which is sufficient for real-time applications.

\end{abstract}
\section{Introduction}
\label{sec:intro}
Speech separation is essential to improve the performance of automatic speech recognition (ASR) under a noisy and speaker-overlapped condition.
Although there have been recent successes in neural-network-based mask estimation \cite{araki2015exploring,drude2017tight,wang2018multi,yoshioka2018recognizing,wang2018spatial} or end-to-end speech separation \cite{von2019all,gu2020enhancing,zhang2020end} for multi-channel signals, beamforming with unsupervised mask estimation is still a powerful speech separation method.
Especially, guided source separation (GSS) \cite{boeddeker2018front}, which involves constructing a beamformer by using diarization information, has performed well on the CHiME-5 corpus \cite{kanda2019guided,zorila2019investigation} that consists of recordings at dinner parties.
GSS was also adopted as a baseline method for the CHiME-6 Challenge \cite{watanabe2020chime} and is still a \textit{de facto} standard for the CHiME-6 corpus \cite{du2020ustc,medennikov2020stc,chen2020ioa}.
A recent study has proven that diarization-first speech separation using GSS is also useful for an ASR system using asynchronous monaural microphones \cite{horiguchi2020utterance}.

There are mainly three advantages of GSS.
The first is that the diarization information can give good initial parameters of a generative model of observations, which makes it possible to work well without pretrained parameters even when there are multiple speakers.
The second advantage is that, although mask estimation is a frequency-wise algorithm, this initialization makes it free from the permutation problem of the frequency domain.
The third advantage is that GSS calculates utterance-wise beamformers so that it works well when a session-level beamformer does not work well, \eg, there is a sampling frequency mismatch between audio channels or speakers are moving around during a session.

The utterance-wise algorithm, however, incurs significant computational cost because several iterations of optimization are required for each utterance.
It also produces latency according to the utterance length.
Thus, this algorithm limits the development of ASR systems based on GSS, especially those of real-time applications.
If GSS is extended to an online algorithm, it is beneficial to implement a highly accurate speech recognition system for an overlapping and conversational speech by combining GSS with online diarization methods for an unrestricted number of speakers \cite{dimitriadis2017developing,zhang2019fully,lin2020speaker} and online ASR \cite{chiu2018monotonic,arivazhagan2019monotonic,fan2019online,moritz2020streaming}.

In this paper, we propose a block-online GSS algorithm to avoid utterance-wise processing.
A block-wise input is processed together with its preceding context to update time-frequency masks and estimate a minimum variance distortionless response (MVDR) beamformers.
There are two benefits of using a preceding context.
One is that the context is helpful to estimate the mask of the current block from only one expectation-maximization (EM) iteration because the context has been processed once in the previous step.
The other is that the context is helpful to solve the frequency permutation problem, the same as with the conventional offline GSS algorithm \cite{boeddeker2018front}.
To reduce the computational cost, the block-wise update only takes into account the parameters of active speakers during the block and its context.
We evaluated the proposed algorithm in both synchronous and asynchronous settings using the CHiME-6 corpus and a meeting corpus recorded using distributed asynchronous microphones.
The experimental results indicate that the proposed algorithm exhibits comparable performance to the conventional offline GSS algorithm with real-time processing.

\section{Related work}
Conventional online mask-based beamforming methods are based on block-wise \cite{higuchi2017online,araki2017online,matsui2018online,togami2019simultaneous} or frame-wise \cite{higuchi2018frame} estimation of time-frequency masks and updating of the beamformer.
There are mainly two approaches for mask estimation: spatial-clustering- and neural-network-based estimation.
Mask estimation based on spatial clustering empirically requires pretrained parameters for initialization \cite{higuchi2017online,araki2017online}, especially when there are multiple speakers, to avoid iterative calculation and the frequency permutation problem.
This is not suitable when the microphone and speaker arrangement is not known in advance.
Mask estimation based on neural networks requires clean training data \cite{matsui2018online,togami2019simultaneous,higuchi2018frame}, which are inaccessible in real conversations.
It is also a problem that such networks typically predetermine the number of input and output channels.
Recently proposed methods accept variable channels of inputs \cite{luo2019fasnet,luo2020endtoend,wang2020neural}, but the number of outputs still has to be known in advance.
The conventional offline GSS and proposed algorithms do not have such limitations.
Recently, Du \etal investigated an online update of the beamformer on the CHiME-6 corpus \cite{du2020ustc}, but the preceding mask estimation based on GSS is an offline algorithm so it cannot work in an online manner.

\section{Overview of guided source separation}
In this section, we explain the conventional offline GSS algorithm \cite{boeddeker2018front}.
We assume that the following procedure is followed for each utterance, as in previous studies \cite{boeddeker2018front,kanda2019guided,zorila2019investigation,watanabe2020chime}.
The GSS involves a complex angular central Gaussian mixture model (cACGMM) \cite{ito2016complex} as its generative model, which has a probability density function at a frequency index $f$ that is determined as
\begin{align}
    p\left(\hat{\vect{x}}_{t,f};\left\{\alpha_f^{(k)},B_{f}^{(k)}\right\}_k\right)&=\sum_k \alpha_f^{(k)}\mathcal{A}\left(\hat{\vect{x}}_{t,f};B_{f}^{(k)}\right),\\
    \hat{\vect{x}}_{t,f}&=\frac{\vect{x}_{t,f}}{\norm{\vect{x}_{t,f}}},
\end{align}
where $\vect{x}_{t,f}\in\mathbb{C}^M$ is the $M$-channel observed signal in a short-time Fourier transform (STFT) domain and $t$ is the time index. The observation can be dereverberated beforehand by using, \eg, the weighted prediction error (WPE) \cite{nakatani2010speech}.
The $\alpha_f^{(k)}$ is the mixture weight of the $k$-th source at $f$, and $\mathcal{A}\left(\vect{x};B\right)$ is a complex angular central Gaussian distribution \cite{kent1997data} parameterized by $B\in\mathbb{C}^{M\times M}$ as
\begin{align}
    \mathcal{A}\left(\vect{x};B\right)\coloneqq\frac{(M-1)!}{2\pi^M\det(B)}\frac{1}{\left(\herm{\vect{x}}\inv{B}\vect{x}\right)^M},
\end{align}
where $\herm{(\cdot)}$ denotes the Hermitian transpose.
The optimization of the cACGMM is done using the EM algorithm. At the E-step, the posterior for each source at a time-frequency bin is calculated as
\begin{align}
    \gamma_{t,f}^{(k)}\leftarrow\frac{\alpha_f^{(k)}\frac{1}{\det\left(B_f^{(k)}\right)}\frac{1}{\left[\herm{\hat{\vect{x}}}_{t,f}\left(B_f^{(k)}\right)^{-1}\hat{\vect{x}}_{t,f}\right]^M}}{\sum_{k'}\alpha_f^{(k')}\frac{1}{\det\left(B_f^{(k')}\right)}\frac{1}{\left[\herm{\hat{\vect{x}}}_{t,f}\left(B_f^{(k')}\right)^{-1}\hat{\vect{x}}_{t,f}\right]^M}}.
    \label{eq:update_gamma_general}
\end{align}
At the M-step the parameters $\alpha_f^{(k)}$ and $B_f^{(k)}$ are updated as follows:
\begin{align}
    \alpha_f^{(k)}&\leftarrow\frac{1}{T}\sum_t\gamma_{t,f}^{(k)},\label{eq:update_alpha}\\
    B_f^{(k)}&\leftarrow M\frac{\sum_t\gamma_{t,f}^{(k)}\frac{\hat{\vect{x}}_{t,f}\herm{\hat{\vect{x}}}_{t,f}}{\herm{\hat{\vect{x}}}_{t,f}\left(B_f^{(k)}\right)^{-1}\hat{\vect{x}}_{t,f}}}{\sum_t\gamma_{t,f}^{(k)}},\label{eq:update_beta_general}.
\end{align}

With GSS, the activities of each speaker are assumed known a priori and used for parameter updates.
Given $d_t^{(k)}\in\{0,1\}$, that is, an activity of source $k$ that takes 1 if the source $k$ is active at $t$ and 0 otherwise, the E-step is replaced with
\begin{align}
    \gamma_{t,f}^{(k)}\leftarrow\frac{\alpha_f^{(k)}d_t^{(k)}\frac{1}{\det\left(B_f^{(k)}\right)}\frac{1}{\left[\herm{\hat{\vect{x}}}_{t,f}\left(B_f^{(k)}\right)^{-1}\hat{\vect{x}}_{t,f}\right]^M}}{\sum_{k'}\alpha_f^{(k')}d_t^{(k')}\frac{1}{\det\left(B_f^{(k')}\right)}\frac{1}{\left[\herm{\hat{\vect{x}}}_{t,f}\left(B_f^{(k')}\right)^{-1}\hat{\vect{x}}_{t,f}\right]^M}}
    \label{eq:update_gamma}
\end{align}
to force the posteriors of inactive sources to be zero.
The diarization information helps to make the model free from the frequency permutation problem because it is frequency-independent.
However, it is still affected by the permutation between the target utterance and noise because the activities of them are always one during the utterance.
To solve this, GSS also uses preceding and subsequent signals of the utterance, which are called ``context,'' for parameter update.
In this paper, we refer to the preceding context as pre-context and subsequent context as post-context.

In the first iteration of the EM updates, $\alpha_f^{(k)}$ and $B_f^{(k)}$ are unknown, so the following \autoref{eq:update_gamma_iter1} and \autoref{eq:update_beta_iter1} are used for the E-step and the update of $B_f^{(k)}$ in the M-step instead, respectively:
\begin{align}
    \gamma_{t,f}^{(k)}&\leftarrow\frac{d_t^{(k)}}{\sum_{k'}d_t^{(k')}}\label{eq:update_gamma_iter1},\\
    B_f^{(k)}&\leftarrow M\frac{\sum_t\gamma_{t,f}^{(k)}\hat{\vect{x}}_{t,f}\herm{\hat{\vect{x}}}_{t,f}}{\sum_t\gamma_{t,f}^{(k)}}\label{eq:update_beta_iter1}.
\end{align}

After convergence, spatial covariance matrices for speech and noise are calculated using the posteriors $\gamma_{t,f}^{(k)}$ as follows:
\begin{align}
    R_{f}^\text{speech}&=\frac{1}{T}\sum_t \gamma_{t,f}^{(k_\text{target})}\vect{x}_{t,f}\herm{\vect{x}}_{t,f}\in\mathbb{C}^{M\times M},\label{eq:covar_speech}\\
    R_{f}^\text{noise}&=\frac{1}{T}\sum_t \left(1-\gamma_{t,f}^{(k_\text{target})}\right)\vect{x}_{t,f}\herm{\vect{x}}_{t,f}\in\mathbb{C}^{M\times M}.\label{eq:covar_noise}
\end{align}
Here we assume that the target source is $k_\text{target}\in\{1,\dots,K\}$.
The MVDR beamformer $\vect{w}_f\in\mathbb{C}^M$ is calculated using the spatial covariance matrices as
\begin{align}
    \vect{w}_f&=\frac{{R_f^\text{noise}}^{-1}R_f^\text{speech}\vect{r}}{\trace{{R_f^\text{noise}}^{-1}R_f^\text{speech}}},\label{eq:beamformer}
\end{align}
where $\vect{r}\in\{0,1\}^M$ is a one-hot vector that corresponds to the reference microphone, which is selected to maximize the signal-to-noise ratio.
Finally, blind analytic normalization \cite{warsitz2007blind} is applied for $\vect{w}_f$ to obtain the final beamformer, which is used for speech enhancement.
The enhanced signal in the STFT domain is calculated as
\begin{align}
    z_{t,f}=\herm{\vect{w}_f}\vect{x}_{t,f}.
\end{align}

\section{Block-online guided source separation}
\label{sec:proposed}
\subsection{Overview}
In the CHiME-6 baseline system, speech separation based on GSS is applied for each utterance.
It takes about 85.44 hours using a single CPU without utterance-wise parallel processing to enhance all the utterances in the development set, which includes about 4.46 hours of recordings.
This processing speed is not sufficient for online processing or for offline ASR systems because it takes over 19x the recording duration for speech separation.

\begin{figure}[t]
    \centering
    \begin{minipage}{0.48\linewidth}
        \includegraphics[width=\linewidth]{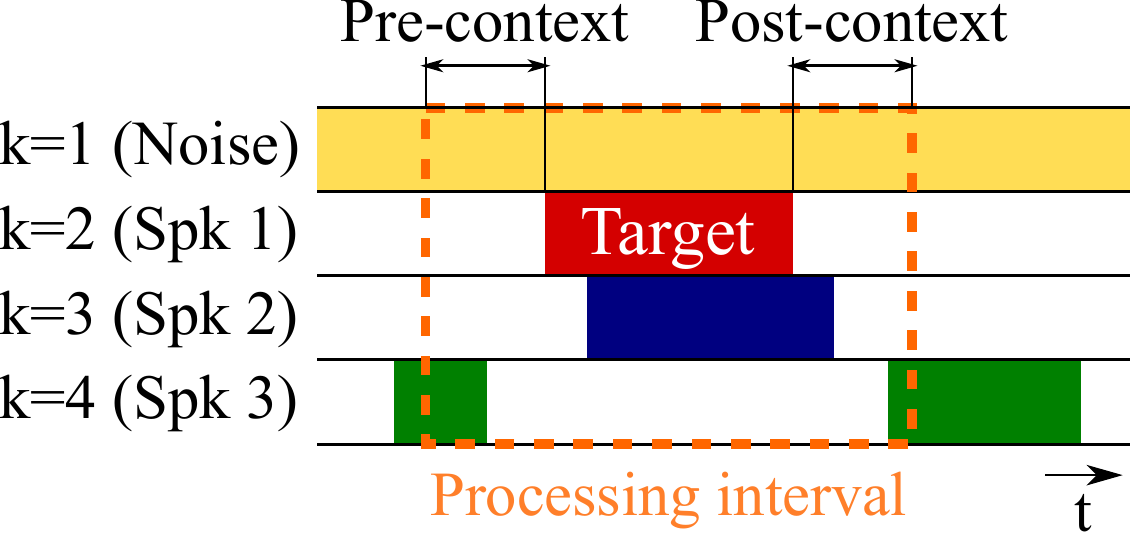}
        \subcaption{Target: Speaker 1}
    \end{minipage}
    \hfill
    \begin{minipage}{0.48\linewidth}
        \includegraphics[width=\linewidth]{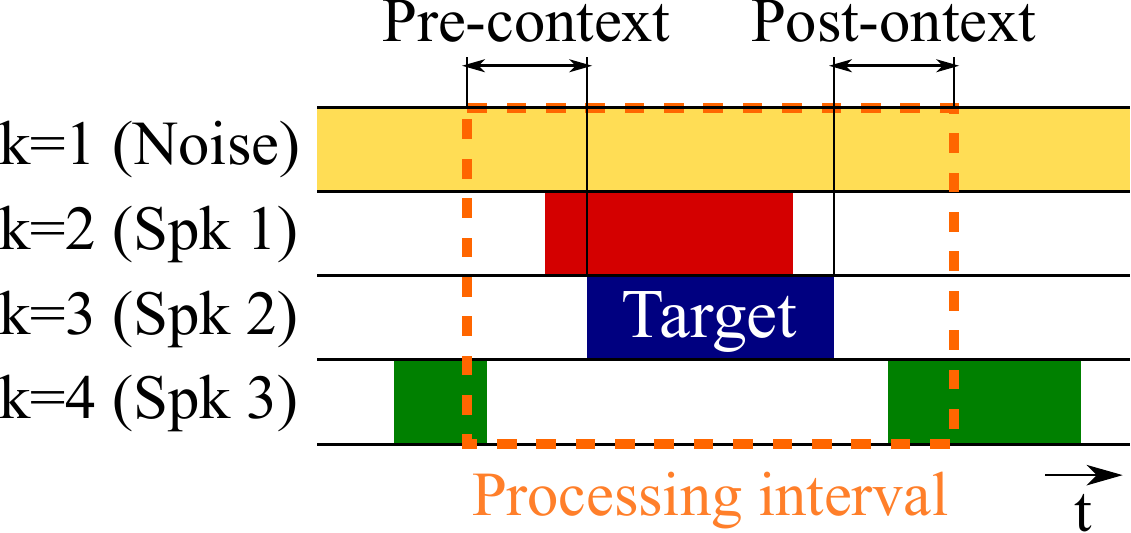}
        \subcaption{Target: Speaker 2}
    \end{minipage}
    \caption{Case when almost same cACGMMs are obtained by conventional utterance-wise offline GSS algorithm.}
    \label{fig:overlap}
\end{figure}

The reason the conventional offline GSS algorithm requires such a long calculation time is the redundancy of the utterance-wise operation. For example, if two speech signals are highly overlapped, as in \autoref{fig:overlap}, the optimized cACGMMs should be almost the same. However, they are optimized independently in the conventional offline GSS algorithm.
It should be also considered in online processing that this algorithm uses a few seconds of signals after each utterance as context. If we use such a post-context even in an online algorithm, it produces latency according to the length of the context.
Moreover, we have to beware that the calculation cost is proportional to the number of speakers.

The proposed algorithm i) updates the parameters of a cACGMM in a block-online manner to avoid redundant calculation as in \autoref{fig:overlap}, ii) only uses a pre-context of each block to reduce latency, and iii) only uses active sources to update parameters to reduce calculation.
Note that the number of speakers in a session does not have to be known a priori because this algorithm determines this adaptively from block-wise input diarization information.

\subsection{Proposed algorithm}
\begin{algorithm}[t]
	\SetAlgoLined
	\DontPrintSemicolon
	\caption{Block-online guided source separation.}
	\label{alg:online_gss}
	\SetAlgoVlined
	\SetKwInOut{Input}{Input}\SetKwInOut{Output}{Input}
	\SetKw{In}{in}
	\SetKw{Continue}{continue}
	\Input{$\left\{\mathbf{X}_n\in\mathbb{C}^{L\times F\times M}\right\}_{n=1}^N$\tcp*{STFT features}}
	\myinput{$\left\{(d_t^{(k)})_{t\in\mathcal{T}_n\atop 1\leq k\leq K_n}\right\}_{n=1}^N$\tcp*{Diarization}}
	\myinput{$C\in\mathbb{Z}_{+}$ \tcp*{\#Pre-context frames}}
	\BlankLine
	$K_0=0$\tcp*{Initial \#Sources}
	\For{$n=1$ \KwTo $N$}{
	    $\mathbf{X}_{n}\leftarrow\mathsf{BlockOnlineWPE}\left(\mathbf{X}_{n}\right)$\;
	    \If(\tcp*[f]{Silent block}){$\sum_{t\in\mathcal{T}_n}\sum_{k=2}^{K_n} d_t^{(k)}=0$}{
	        \Continue
	    }
    	$\mathcal{K}\leftarrow\left\{k\relmid1\leq k\leq K_n,\sum_{t\in\mathcal{T}_n^+}d_t^{(k)}>0\right\}$\tcp*{Set of active sources}
    	\ForEach{$f\in\{1,\dots,F\}$}{
    	    \For(\tcp*[f]{New sources}){$k=K_{n-1}+1$ \KwTo $K_n$}{
        	    $\gamma_{t,f}^{(k)}\leftarrow 0$ for $t\in\mathcal{T}_n^c$\;
        	    $\Gamma_f^{(k)}\leftarrow 0$\;
        	    $B_f^{(k)}\leftarrow O_M$\;
    	    }
    	    Initialize $\gamma_{t,f}^{(k)}$ for $(t,k)\in\mathcal{T}_n\times\mathcal{K}$ by \autoref{eq:update_gamma_iter1}\;
        	Update $\alpha_f^{(k)}$ for $k\in\mathcal{K}$ using $\hat{\mathbf{X}}_n^+$ by \autoref{eq:update_alpha}\;
        	Calculate $B_{n,f}^{+(k)}$ for $k\in\mathcal{K}$ using $\hat{\mathbf{X}}_n^+$ by \autoref{eq:update_beta_current_block}\;
        	Update $B_f^{(k)}$ for $k\in\mathcal{K}$ using $\hat{\mathbf{X}}_n^+$ by \autoref{eq:update_beta_accumulate}--\autoref{eq:update_accumulative_gamma} or \autoref{eq:update_beta_decay}\;
        	Update $\gamma_{t,f}^{(k)}$ for $(t,k)\in\mathcal{T}_n^+\times\mathcal{K}$ by \autoref{eq:update_gamma}
	    }
        \ForEach{utterance spoken during $\mathcal{T}_n$}{
            \tcc{assume that the utterance started at $t_s$ and ended at $t_e$}
            Calculate $R_f^\text{speech}$ and $R_f^\text{noise}$ for $f\in\{1,\dots,F\}$ from $\mathbf{X}_n^+$ by \autoref{eq:covar_speech}--\autoref{eq:covar_noise}\;
            Calculate $\vect{w}_f$ for $f\in\{1,\dots,F\}$ by \autoref{eq:beamformer}\;
            Output an enhanced audio $\left(\herm{\vect{w}_f}\vect{x}_{t,f}\right)_{\max(t_s,(n-1)L+1)\leq t\leq \min(t_e,nL)\atop f\in\{1,\dots,F\}}$
        }
	}
\end{algorithm}

The proposed algorithm is shown in \autoref{alg:online_gss}.
Let $L$ be the length of a block along the time axis, $C$ be the length of a pre-context of a block, $N$ be the length of a sequence of blocks, and $K_n (K_1\leq K_2\leq\dots\leq K_N$) be the number of sources appearing no later than the $n$-th block.
For sake of simplicity, we define the set of time indexes in the $n$-th block as $\mathcal{T}_n\coloneqq\{(n-1)L+1,\dots,nL\}$, that in the pre-context of the $n$-th block as $\mathcal{T}_n^c\coloneqq\{(n-1)L-C+1,\dots,(n-1)L\}\cap\mathbb{N}$, and the union of them as $\mathcal{T}_n^+\coloneqq\mathcal{T}_n\cup\mathcal{T}_n^c$.
To process the $n$-th block, we use samples in the previous blocks as the pre-context. Thus, we prepared a $C$-length queue and store the most recent $C$ frames to use them in the block-online processing.

We assume that inputs are blocked STFT features $\left\{\mathbf{X}_n\right\}_{n=1}^N$, where $\mathbf{X}_n=\left(\vect{x}_{t,f}\right)_{t\in\mathcal{T}_n\atop f\in\{1,\dots,F\}}\in\mathbb{C}^{L\times F\times M}$, and their corresponding diarization results $\left\{(d_t^{(k)})_{t\in\mathcal{T}_n\atop 1\leq k\leq K_n}\right\}_{n=1}^N$.
Note that $k=1$ corresponds to noise, whose activities $d_t^{(1)}$ are always one, and $k\geq 2$ corresponds to speakers.
First, the initial number of sources $K_0$ is set to zero (Line 1 in \autoref{alg:online_gss}; L1).
For each block (L2), the block-online WPE \cite{drude2018nara} is applied for dereverberation of the input features (L3).
If there is no active speaker in the block, we finish processing for the input block (L4--5).
If active speakers exist in the block, we extract the set of active sources $\mathcal{K}$ in the block and its context (L6).

The parameters of cACGMM are then updated for each frequency index $f$ using the block and its pre-context (L7).
To calculate the posteriors using \autoref{eq:update_gamma}, the mixture weight $\alpha_f^{(k)}$ and the matrix parameter $B_f^{(k)}$ for each active speaker are required.
However, we do not have such $\alpha_f^{(k)}$ because the active source set $\mathcal{K}$ differs among blocks, and we also do not have reliable $B_f^{(k)}$ for new speakers.
Therefore, in this online strategy, we first update $\alpha_f^{(k)}$ and $B_f^{(k)}$ using initial estimations of $\gamma_{t,f}^{(k)}$ calculated from the input diarization information, and then estimate $\gamma_{t,f}^{(k)}$ using the estimated $\alpha_f^{(k)}$ and $B_f^{(k)}$.

For each new speaker (L8), we set the posteriors $\gamma_{t,f}^{(k)}$ during the context by zero (L9) because the new speakers are not active during the context interval.
The value for the accumulation of posteriors $\Gamma_f^{(k)}$ is also initialized with zero (L10) and the matrix parameter $B_f^{(k)}$ is initialized with $M\times M$ zero matrix $O_M$ (L11).
We also initialize the posteriors of all the active speakers during the input block by \autoref{eq:update_gamma_iter1} (L12).
The mixture weight $\alpha_f^{(k)}$ for each source is then updated (L13) using \autoref{eq:update_alpha}.
As described above, the sets of speakers differ from block to block; thus $\alpha_f^{(k)}$ is updated by \autoref{eq:update_alpha} without any smoothing over blocks.
On the other hand, in this study, we used two update strategies to update the matrix parameter $B_f^{(k)}$ (L14--15).
One is the \textbf{accumulation} strategy, which updates $B_f^{(k)}$ to be closer to the offline estimation.
We first calculate the matrix parameters using  $\hat{\mathbf{X}}_n^+\coloneqq\left(\hat{\vect{x}}_{t,f}\right)_{t\in\mathcal{T}_n^+\atop f\in\{1,\dots,F\}}$ by
\begin{align}
    B_{n,f}^{+(k)}=\begin{cases} M\frac{\sum_{t\in\mathcal{T}_n^+}\gamma_{t,f}^{(k)}\frac{\hat{\vect{x}}_{t,f}\herm{\hat{\vect{x}}}_{t,f}}{\herm{\hat{\vect{x}}}_{t,f}\left(B_f^{(k)}\right)^{-1}\hat{\vect{x}}_{t,f}}}{\sum_{t\in\mathcal{T}_n^+}\gamma_{t,f}^{(k)}}&(k\leq K_{n-1})\\
    M\frac{\sum_{t\in\mathcal{T}_n^+}\gamma_{t,f}^{(k)}\hat{\vect{x}}_{t,f}\herm{\hat{\vect{x}}}_{t,f}}{\sum_{t\in\mathcal{T}_n^+}\gamma_{t,f}^{(k)}}&(k> K_{n-1})
    \end{cases}.\label{eq:update_beta_current_block}
\end{align}
By using this, $B_f^{(k)}$ is updated by
\begin{align}
    B_f^{(k)}&\leftarrow\frac{\Gamma_f^{(k)}}{\Gamma_f^{(k)}+\sum_{t\in\mathcal{T}_n}\gamma_{t,f}^{(k)}}B_f^{(k)}+\frac{\sum_{t\in\mathcal{T}_n}\gamma_{t,f}^{(k)}}{\Gamma_f^{(k)}+\sum_{t\in\mathcal{T}_n}\gamma_{t,f}^{(k)}}B_{n,f}^{+(k)},\label{eq:update_beta_accumulate}
\end{align}
where $\Gamma_f^{(k)}$ is an accumulation of the posteriors, which is updated after updating $B_f^{(k)}$ as follows:
\begin{align}
    \Gamma_{f}^{(k)}\leftarrow\Gamma_{f}^{(k)}+\sum_{t\in\mathcal{T}_n}{\gamma_{t,f}^{(k)}}.
    \label{eq:update_accumulative_gamma}
\end{align}
The accumulation strategy is known to be effective when there is a beamformer that works well through a session \cite{higuchi2017online,matsui2018online}. However, if there is a sampling frequency mismatch between audio channels or speakers are moving around during a session, such a session-wise beamformer is not sufficient and its block-level refinement is required for performance improvement \cite{araki2017meeting,araki2018meeting}.
In such situations, the parameters of cACGMM should be updated to have temporal locality.
Therefore, we also used the \textbf{decay} strategy, in which we update $B_f^{(k)}$ as follows:
\begin{align}
    B_f^{(k)}&\leftarrow\eta B_f^{(k)}+ B_{n,f}^{+(k)},\label{eq:update_beta_decay}
\end{align}
where $\eta\in\left[0,1\right)$ is a factor of decay.
The posteriors $\gamma_{t,f}$ during $\mathcal{T}_n^+$ are then updated using \autoref{eq:update_gamma} (L16).

The updates of $\alpha_f^{(k)}$, $B_f^{(k)}$, and $\gamma_{t,f}^{(k)}$ above are conducted using the samples of the block and its pre-context.
By using the pre-context, the permutation problem can be solved, as in the conventional offline GSS algorithm.
Furthermore, the posteriors $\gamma_{t,f}^{(k)}$ for the context are computed once in the previous iteration; they are helpful for accurate estimation of the mixture weights $\alpha_f^{(k)}$ and matrix parameter $B_f^{(k)}$, and eventually calculation of posteriors for the current block only from one EM iteration.

After the update of the cACGMM, an MVDR beamformer is calculated using the optimized cACGMM.
For each utterance spoken during the block, regardless of whether the utterance is finished, we calculate spatial covariance matrices from $\mathbf{X}_n^+$, calculate a beamformer, and output enhanced audio during the utterance (L17--20).

\section{Experiments}
\subsection{Experimental settings}
\begin{figure*}[t!]
    \begin{minipage}[c]{.60\linewidth}
        \centering
        \tblcaption{Evaluation Corpora.}
        \label{tbl:datasets}
        \begin{tabular}{@{}lcccccc@{}}
        \toprule
        Corpus&Session&\#Mic&Duration&\#Spk&\#Utt&Overlap\\\midrule
        \multirow{2}{*}{CHiME-6 dev \cite{watanabe2020chime}}&S02&12&2:28:22&4&3822&\SI{52.9}{\percent}\\
        &S09&10&1:59:20&4&3615&\SI{45.8}{\percent}\\\midrule
        \multirow{8}{*}{Meeting \cite{horiguchi2020utterance}}&I&2/3/6/11&19:49&7&160&\SI{6.9}{\percent}\\
        &I\hspace{-.1em}I&2/3/6/11&14:27&8&150&\SI{14.0}{\percent}\\
        &I\hspace{-.1em}I\hspace{-.1em}I&2/3/6/11&13:13&5&198&\SI{16.6}{\percent}\\
        &I\hspace{-.1em}V&2/3/6/11&12:08&7&184&\SI{19.9}{\percent}\\
        &V&2/3/6/11&12:37&6&80&\SI{5.5}{\percent}\\
        &V\hspace{-.1em}I&2/3/6/11&16:50&6&256&\SI{14.7}{\percent}\\
        &V\hspace{-.1em}I\hspace{-.1em}I&2/3/6/11&16:25&7&223&\SI{11.3}{\percent}\\
        &V\hspace{-.1em}I\hspace{-.1em}I\hspace{-.1em}I&2/3/6/11&11:25&7&185&\SI{19.9}{\percent}\\
        \bottomrule
        \end{tabular}
    \end{minipage}
    \hfill
    \begin{minipage}[c]{.36\linewidth}
        \centering
        \includegraphics[width=\linewidth]{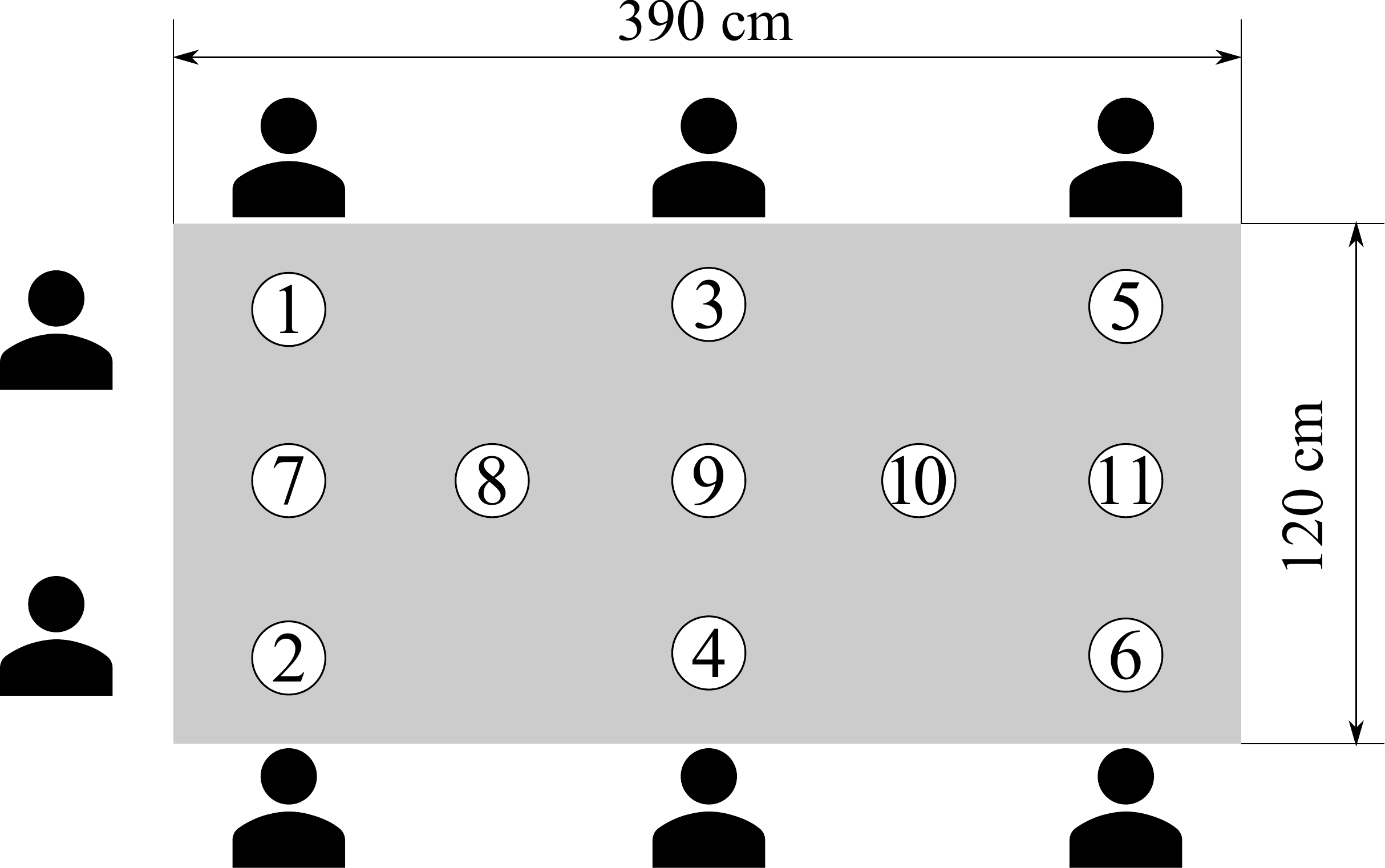}
        \caption{Recording environment of meeting corpus. 11 smartphones, each equipped with monaural microphone, were distributed on table.}
        \label{fig:recording_environment}
    \end{minipage}
\end{figure*}
To evaluate the proposed algorithm, we used two corpora.
The statistics of the two corpora are listed in Table \ref{tbl:datasets}.

One is the CHiME-6 corpus \cite{watanabe2020chime}, of which each session consists of a dinner party scenario with four participants. It was recorded using six distributed Kinect\textsuperscript{\textregistered} v2 devices, each equipped with four microphones. The sampling frequency mismatch between devices was manually corrected. However, participants were moving around the kitchen, dining room, and living room; thus, a fixed beamformer for each participant does not exist.
We evaluated the word error rates (WERs) of two sessions (S02 \& S09) in the development set.
It is because clean speech is not available in the corpus; therefore, we cannot use an evaluation metric such as the signal-to-distortion ratio (SDR) to directly calculate speech separation accuracy.
Following the CHiME-6 baseline system provided as a Kaldi recipe\footnote{\url{https://github.com/kaldi-asr/kaldi/tree/master/egs/chime6/s5_track1}}, we used 12-channel signals, \ie, the outer two microphones of each device, for S02 and 10-channel signals for S09 because the recording of the fifth device was unavailable.
For a speech recognition module, we used an acoustic model based on a factorized time delay neural network and 3-gram language model with two-stage decoding as in the baseline system.

The other is a meeting corpus \cite{horiguchi2020utterance}, which was recorded using 11 asynchronous distributed microphones, as shown in \autoref{fig:recording_environment}.
This corpus contains eight sessions of a Japanese meeting each with 5-8 participants, and the sampling frequency mismatch between microphones are not corrected.
We evaluated the character error rates (CERs) using various combinations of microphones: 2 microphones (\numcircle{8}\&\numcircle{10}), 3 microphones (\numcircle{7}\&\numcircle{9}\&\numcircle{11}), 6 microphones (\numcircle{1}-\numcircle{6}), and 11 microphones (\numcircle{1}--\numcircle{11}).
We used our meeting transcription system \cite{horiguchi2020utterance} for evaluation by replacing its speech enhancement module with the proposed online GSS.

A parameter set for the proposed algorithm is shown in \autoref{tbl:parameters}.
To enable real-time processing, the tap size of the WPE was set to two, which was set to 10 in the offline baselines \cite{watanabe2020chime,horiguchi2020utterance}. 
The block size $L$ and the pre-context size $C$ were varied among the experiments.
From the view point of utterance, the number of pre-context frames for each utterance $c_\text{pre}$ fulfills $C\leq c_\text{pre}\leq C+L-1$ and that of post-context frames for each utterance $c_\text{post}$ fulfills $0\leq c_\text{post}\leq L-1$.
The offline baseline uses \SI{10}{\second} of pre- and post-contexts for the CHiME-6 corpus and \SI{15}{\second} of them for the meeting corpus, as in previous studies \cite{watanabe2020chime,horiguchi2020utterance}.
For diarization information $d_t^{(k)}$, we used oracle speech segments.
In the CHiME-6 evaluation, we also used estimated diarization results obtained by a single iteration of target-speaker voice activity detection (TS-VAD) \cite{medennikov2020targetspeaker}\footnote{\url{https://github.com/kaldi-asr/kaldi/tree/master/egs/chime6/s5b_track2}}.

Note that a block-wise input sometimes contains a new speaker $k$ with a very limited number of active frames.
In such a case, the estimated matrix parameter $B_f^{(k)}$ is not reliable.
To avoid using such an unreliable parameter to process the next block, we treated the speaker $k$ as a new speaker in the next block processing, \ie, we conducted L9--11 in \autoref{alg:online_gss} once again for the speaker $k$, if the duration of the active frames was less than \SI{0.2}{\second}.

\begin{table}[t]
    \centering
    \caption{Parameters used in online experiments.}
    \begin{tabular}{@{}lc@{}}
        \toprule
        Audio sampling rate & \SI{16}{\kHz}\\
        STFT window length & \SI{64}{\ms}\\
        STFT window shift & \SI{16}{\ms}\\
        STFT window function & Hanning\\
        WPE taps & 2 frames\\
        WPE delay & 2 frames\\
        WPE decay factor & 0.9\\
        \bottomrule
    \end{tabular}
    \label{tbl:parameters}
\end{table}

\subsection{Results}
\label{sec:format}
We first evaluated the performance of the proposed online GSS algorithm on the CHiME-6 development set using the oracle segments.
The results are shown in \autoref{tbl:chime6}.
The decay strategy always performed better than the accumulation strategy when the same parameters $(L,C)$ were used.
This indicates that the decay strategy is suitable for home environments in which speakers are moving during a session.
With the decay strategy, the WERs improved by using the pre-context from \SI{57.3}{\percent} to \SI{51.6}{\percent} when $L=150$ (\SI{2.4}{\second}) and from \SI{56.0}{\percent} to \SI{52.3}{\percent} when $L=300$ (\SI{4.8}{\second}).
By comparing the results of $(L,C)=(300,0)$ and $(150,150)$, we can also observe that the pre-context improved WERs from \SI{56.0}{\percent} to \SI{51.6}{\percent} even if the lengths of the processing unit $L+C$ are the same.
These results indicate that using pre-context for parameter update is important for accurate mask estimation only from one EM iteration.

We also evaluated the performance of the proposed online GSS algorithm using estimated diarization results.
They were obtained by a single iteration of TS-VAD, which showed a diarization error rates of \SI{46.5}{\percent} and \SI{53.62}{\percent} on S02 and S09, respectively.
The results shown in \autoref{tbl:chime6_tsvad} indicate that the proposed method works as well as the offline GSS, even if it is based on the estimated diarization results.

\begin{table}[t]
    \centering
    \caption{WERs (\%) on CHiME-6 development set.}
    \centering
    \subcaption{With oracle segments.}
    \label{tbl:chime6}
    \scalebox{1.0}{
    \resizebox{\linewidth}{!}{
    \begin{tabular}{@{}lccccc@{}}
        \toprule
        &&&\multicolumn{3}{c}{Session}\\\cmidrule(l){4-6}
        Algorithm&Block $L$&Context $C$&S02&S09&All\\\midrule
        Offline \cite{boeddeker2018front,watanabe2020chime}&---&---&52.2&51.1&51.8\\
        Online (Accumulation)&150&0&59.4&64.7&61.5\\
        Online (Accumulation)&300&0&59.1&64.3&61.1\\
        Online (Accumulation)&150&150&62.9&63.6&63.1\\
        Online (Accumulation)&300&300&62.8&63.8&63.2\\
        Online (Decay, $\eta=0.9$)&150&0&55.7&59.9&57.3\\
        Online (Decay, $\eta=0.9$)&300&0&54.1&59.0&56.0\\
        Online (Decay, $\eta=0.9$)&150&150&50.6&53.3&51.6\\
        Online (Decay, $\eta=0.9$)&300&300&51.4&53.7&52.3\\
        \bottomrule
    \end{tabular}
    }
    }\\
    \par\bigskip
    \subcaption{With estimated diarization results obtained by TS-VAD.}
    \label{tbl:chime6_tsvad}
    \scalebox{1.0}{
    \resizebox{\linewidth}{!}{
    \begin{tabular}{@{}lccccc@{}}
        \toprule
        &&&\multicolumn{3}{c}{Session}\\\cmidrule(l){4-6}
        Algorithm&Block $L$&Context $C$&S02&S09&All\\\midrule
        Offline \cite{boeddeker2018front,watanabe2020chime}&---&---&70.2&70.0&70.1\\
        Online (Decay, $\eta=0.9$)&150&150&70.1&71.3&70.6\\
        \bottomrule
    \end{tabular}
    }
    }
\end{table}

\begin{figure}[t]
    \centering
    \includegraphics[width=\linewidth]{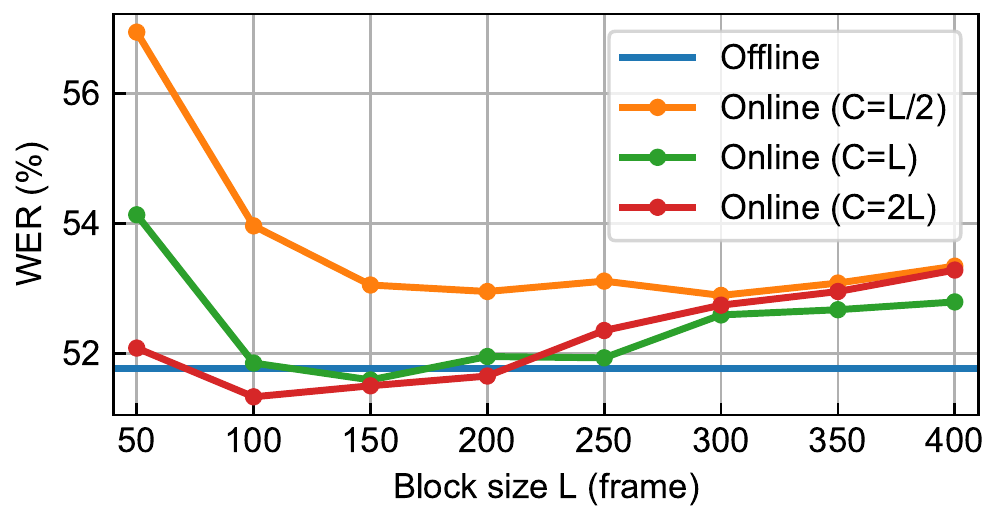}
    \caption{WERs (\%) on CHiME-6 development set with various block size $L$ and pre-context size $C$. Decay strategy with $\eta=0.9$ was used for proposed algorithm.}
    \label{fig:chime6_lc}
\end{figure}

\begin{table*}[t]
    \centering
    \caption{CERs (\%) on meeting corpus recorded using asynchronous distributed microphones. The block size $L$ and context size $C$ were set to 150 for proposed algorithm.}
    \label{tbl:meeting}
    \begin{tabular}{@{}clccccccccc@{}}
        \toprule
        &&\multicolumn{8}{c}{Session}\\\cmidrule(l){3-10}
        \#Mic&Method&I&I\hspace{-.1em}I&I\hspace{-.1em}I\hspace{-.1em}I&I\hspace{-.1em}V&V&V\hspace{-.1em}I&V\hspace{-.1em}I\hspace{-.1em}I&V\hspace{-.1em}I\hspace{-.1em}I\hspace{-.1em}I&All\\\midrule
        &Offline&19.4&21.5&26.8&31.0&19.3&34.9&32.2&40.7&27.1\\
        2&Online (Accumulation)&19.9&25.7&27.9&32.1&20.0&36.7&32.2&40.2&28.2\\
        &Online (Decay, $\eta=0.9$)&20.3&20.8&27.3&31.5&19.7&36.2&32.3&39.9&27.4\\\midrule
        &Offline&21.7&22.0&31.4&29.3&20.0&38.2&36.0&41.0&28.9\\
        3&Online (Accumulation)&21.9&23.4&34.2&31.3&20.5&39.6&38.9&43.4&30.4\\
        &Online (Decay, $\eta=0.9$)&21.1&22.9&32.9&30.6&19.6&37.6&36.3&42.0&29.2\\\midrule
        &Offline&19.3&18.2&24.7&23.5&17.7&28.4&28.3&34.1&23.5\\
        6&Online (Accumulation)&18.6&20.1&24.1&24.2&17.7&31.8&31.0&37.0&24.6\\
        &Online (Decay, $\eta=0.9$)&18.7&17.9&25.6&23.8&17.9&32.2&31.9&37.2&24.6\\\midrule
        &Offline&17.0&16.3&21.7&21.2&17.7&27.0&27.0&32.8&21.8\\
        11&Online (Accumulation)&17.2&16.9&20.3&22.9&17.6&26.2&27.8&33.2&21.9\\
        &Online (Decay, $\eta=0.9$)&17.1&15.4&21.7&21.8&17.5&27.0&26.7&33.4&21.7\\
        \bottomrule
    \end{tabular}
\end{table*}

\autoref{fig:chime6_lc} shows WERs with various block sizes $L$ and pre-context sizes $C$ using the oracle segments.
The decay strategy with $\eta=0.9$ was used for the proposed online GSS algorithm.
We found that the WERs slightly degraded as $L$ became larger.
This means that the MVDR beamformer should be calculated at short intervals under speaker-moving conditions, such as in CHiME-6.
We also observed that the WERs are highly dependent on $C$ when $L$ is small, \ie, $L=50$.
This indicates that the size of a unit to update parameters and calculate MVDR beamformers should be large to some extent.
In this case, $L+C\geq200$ was sufficient to avoid performance degradation due to the smallness of the unit.

We also evaluated the proposed algorithm on the meeting corpus using various combinations of asynchronous distributed microphones.
In this experiment, the block size $L$ and the pre-context size $C$ were set to 150 frames and the oracle segments were used.
The results are shown in \autoref{tbl:meeting}.
The proposed algorithm performed comparatively with the conventional offline GSS, \eg, \SI{21.8}{\percent} CER with the offline GSS and \SI{21.7}{\percent} CER with the online GSS with the decay strategy, respectively, by using 11 microphones.
In terms of the update strategy of the matrix parameter $B_f^{(k)}$, the decay strategy showed equivalent or better CERs than the accumulation strategy in overall performance, which is the same trend as the results from the CHiME-6 corpus.

Finally, we show the execution times on the CHiME-6 development set in \autoref{tbl:exec_time}. We showed the average and standard deviation of ten trials using Intel\textsuperscript{\textregistered} Xeon\textsuperscript{\textregistered} Gold 6132 CPU@\SI{2.60}{\GHz} with a single thread.
The proposed algorithm enhanced all utterances in the CHiME-6 development set within \SI{2.65}{hours}, while the conventional offline GSS algorithm required \SI{85.44}{hours} by using the Kaldi CHiME-6 recipe. This means that the proposed algorithm is about 32x faster than the conventional offline GSS algorithm on the CHiME-6 corpus.
In terms of real-time processing, the proposed algorithm skips most of the processing (L6--20 in \autoref{alg:online_gss}) if there is no speech activity in the input block; thus, the execution time should be compared with speech duration.
As shown in \autoref{tbl:exec_time}, the execution time for each session is less than the speech duration; so we can fairly conclude that the proposed algorithm can be used in real-time applications.

\begin{table}[t]
    \centering
    \caption{Execution times on CHiME-6 development set. Mean and standard deviation among 10 trials are shown. The block size $L$ and the pre-context size $C$ were set to $150$, which corresponds to \SI{2.4}{\second}.}
    \label{tbl:exec_time}
    \scalebox{0.93}{
    \begin{tabular}{@{}cccccc@{}}
        \toprule
        &&\multicolumn{2}{c}{Duration (s)}&\multicolumn{2}{c}{Execution time (s)}\\\cmidrule(l){3-4}\cmidrule(l){5-6}
        Session&\#Mic&Total&Speech&Offline&Online\\\midrule
        S02& 12& 8902&8492&$183529\pm9567$&$6135\pm93$\\
        S09& 10& 7160&5552&$124054\pm7114$&$3418\pm66$\\\bottomrule
    \end{tabular}
    }
\end{table}

\section{Conclusion}
In this paper, we proposed a block-online algorithm of GSS.
A block-wise input and its pre-context are used together to update the cACGMM parameters and posteriors of active speakers, which are then used to calculate MVDR beamformers.
The proposed algorithm achieved almost the same performance as the conventional offline GSS algorithm on both the CHiME-6 and the meeting corpora, but with 32x faster calculation, which is sufficient for real-time processing.
Future work will involve a simultaneous evaluation of the proposed online GSS algorithm and online diarization \cite{dimitriadis2017developing,zhang2019fully,lin2020speaker,xue2021online}.

\bibliographystyle{IEEEbib}
\bibliography{mybib}

\end{document}